\documentclass[fleqn,usenatbib]{mnras}

\usepackage{newtxtext,newtxmath}

\usepackage[T1]{fontenc}
\usepackage{ae,aecompl}

\usepackage{graphicx}	
\usepackage{amsmath}	

\newcommand{\chandra}{{\it Chandra}}
\newcommand{\nustar}{{\it NuSTAR}}
\newcommand{\suzaku}{{\it Suzaku}}
\newcommand{\swift}{{\it Swift}}
\newcommand{\xmm}{{\it XMM-Newton}}

\newcommand{\red}{\textcolor{black}}
\newcommand{\ergps}{erg~cm$^{-2}$~s$^{-1}$}

\newcommand{\src}{Arp~302N}
\newcommand{\srcfull}{Arp~302}

\title[X-rays of \srcfull]{The origin of the X-ray emission from the non-starburst gas-rich luminous infrared galaxies \srcfull}

\author[J. Jiang et al.]{
Jiachen Jiang,$^{1}$\thanks{E-mail: jj447@cam.ac.uk}
 William Baker,$^{2,3}$Andrew Young$^{4}$ and Luigi Gallo$^{5}$
\\
$^{1}$Institute of Astronomy, University of Cambridge, Madingley Road, Cambridge CB3 0HA, UK\\
$^{2}$Cavendish Laboratory – Astrophysics Group, University of Cambridge, 19 JJ Thomson Avenue, Cambridge CB3 0HE, UK\\
$^{3}$Kavli Institute for Cosmology, University of Cambridge, Madingley Road, Cambridge CB3 0HA, UK\\
$^{4}$School of Physics, Tyndall Avenue, University of Bristol, Bristol BS8 1TH, UK\\
$^{5}$Department of Astronomy and Physics, Saint Mary’s University, 923 Robie Street, Halifax, NS, B3H 3C3, Canada
}

\date{Accepted XXX. Received YYY; in original form ZZZ}

\pubyear{2022}

\begin{document}
\label{firstpage}
\pagerange{\pageref{firstpage}--\pageref{lastpage}}
\maketitle

\begin{abstract}
\red{We present an analysis of the \xmm\ observation of luminous infrared merging galaxies \srcfull\ and a joint re-analysis of its \chandra\ observation. In particular, we focus on the more significant X-ray} emitter of the pair, \src. \chandra\ detects significant soft X-ray emission from the hot gas in the star-forming region of \src\ spreading up to 12\,kpc. We estimate the star-formation rate of \src\ to be around 1-2 $M_{\odot}$ yr$^{-1}$ based on the X-ray luminosity of the star-forming region, similar to previous measurements at longer wavelengths. \red{\chandra\ and \xmm\ observations show evidence of a Si\,\textsc{xiii} emission line with 86\% confidence. Our best-fit model infers a super-solar silicon abundance in the star-forming region, likely related to the past core-collapse supernovae in this galaxy. Similar silicon overabundance was reported in the circumstellar medium of core-collapse supernova remnants in our Galaxy.}

\red{We also detect narrow Fe K$\alpha$ and Fe K$\beta$ (98.6\% confidence) emission lines as part of the AGN emission. Our best-fit spectral model using \texttt{mytorus} indicates the evidence of a heavily obscured power-law emission with $N_{\rm H}>3\times10^{24}$\,cm$^{-2}$ in addition to a weak, unobscured power-law emission.  The scattering fraction of the unobscured power-law emission from Compton-thin materials is 0.7\%. All these spectral features suggest evidence of a Seyfert 2-like AGN in \src. The X-ray measurement of its AGN activity is consistent with the previous \textit{Spitzer} measurement of the same object.}

\end{abstract}




\section{Introduction}

Luminous infrared galaxies (LIRGs) have a high infrared luminosity of $L_{\rm IR, 8-1000 \mu m} > 10^{11} L_{\odot}$ \red{ \citep[e.g.,][]{perez21}. Many of them are believed to be powered by star-formation and/or an active galactic nucleus \citep[AGN, e.g.][]{sanders96,perez21}.}  \red{Obervations and theoretical studies suggest that mergers may be the driving mechanism, lead to star formation and possibly fuel the AGN \citep[e.g.,][]{clements96,hopkins06,yuan10,kawaguchi20,ricci23}. The likely mechanism for the enhanced infrared luminosity is massive stars and/or AGN emitting ultraviolet photons, which heat a large amount of dust and gas in the galaxy, resulting in the strong infrared emission \citep{stierwalt13}. Therefore, gas-rich galaxy mergers are considered a key stepping stone to the understanding of the coevolution of galaxies and their central supermassive black holes \citep[SMBHs,][]{hopkins06,zhuang2023}. A sustainable supply of gas-rich galaxy mergers would fuel their galaxy and SMBH coevolution, whereby giant molecular clouds could get funnelled towards the central region by the effects of dynamical friction \citep[boosting growth,][]{Lin_2023}{}{}, whilst AGN activity caused by the supply of gas regulates the process \citep{liao2023}.}

A specific subgroup of infrared galaxies is called ultra-luminous infrared galaxies (ULIRGs) and have an even higher luminosity $L_{\rm IR} > 10^{12} L_{\odot}$. \red{These could likely have once been ordinary LIRGs boosted by mergers \citep{stierwalt13}. 
In addition, recent evidence has been found for a dual AGN fraction of at least 5\% in a sample of local ULIRGs \citep{efstathiou22}.}

\red{The study of the circumnuclear environments of the accreting SMBHs in U/LIRGs is difficult due to gas and dust obscuration. X-rays, especially hard X-rays less affected by obscuration, become a powerful tool to probe their hidden AGNs. For instance, \citet{yamada21} leveraged all archival X-ray data from \chandra, \xmm, \suzaku, \nustar\ and \swift\ of the \swift-BAT LIRGs and ULIRGs and found a high fraction of Compton-thick absorption in both late and early mergers, 25\% and 65\% respectively. A similar conclusion was drawn by an earlier piece of work in \citet{ricci17}.} The young population of the binary system, e.g., high-mass X-ray binaries (HMXBs) and supernova remnants, O-type stars in the star-forming region of these infrared galaxies, also produce a lot of emission in the soft X-ray band. Emission from the hot gas of temperature around 0.1--1\,keV is often seen in the soft X-ray band. This hot gas is believed to be heated by shock fronts generated by supernova explosions and stellar winds \citep{persic04}. 

\red{Leveraging the $0.5''$ resolution of \chandra\ ACIS, the Great Observatory All-sky LIRG Survey (GOALS) also studied the X-ray surface brightness distribution of U/LIRGs. Extended soft X-ray emission is clear even by visual inspection in most LIRGs. Most of GOALs U/LIRGs' soft X-ray emission, e.g., in the 0.5--2\,keV band, spreads up to 2--10\,kpc with a median value of 5.3\,kpc \citep[][]{iwasawa11}. A more compact, point-like hard X-ray emission is often found, e.g., within 1--3\,kpc, in the GOALS sample. Exceptions were occasionally seen in a small number of objects where soft and hard X-ray emission regions are compact within a couple of kpc \citep{iwasawa11}.}


In this work, we study an infrared galaxy pair \srcfull\ in the X-ray band. A \swift\ ultraviolet image of \srcfull\ (\src\ and Arp 302S) is shown in Fig.\,\ref{pic_om}. \src\ is an edge-on galaxy and Arp 302S is a face-on galaxy \citep{armus09}. \srcfull\ is a luminous infrared source with $L_{\rm IR}=4.2\times10^{11}L_{\odot}$. A burst of star formation usually explains the large infrared luminosity and almost all luminous infrared galaxies like \srcfull\ are found to be rich in molecular gas \citep[e.g.][]{sanders91,solomon92}. Interestingly, most of the star-forming activity in local LIRGs happens in a small region within a few kpc \citep{strickland07}. However, the CO emission extends over 23\,kpc in \src\ and 10\,kpc in Arp 302S, contrasting starbursts. Together with the low star-forming efficiency measured by $L_{\rm IR}/M{\rm (H_{2})}$, \citet{lo97} concluded that the high infrared luminosity is not due to a high star-formation rate (SFR) but an unusual amount of molecular gas in \srcfull. Stars are forming a rate of only 1-4 $M_{\odot}$ yr$^{-1}$ in \srcfull\ \citep{lo97}. 

\red{Joining the forces of \textit{Spitzer}, \textit{GALEX} and \textit{HST} and \textit{Chandra}, \citet{armus09} conducted a multi-wavelength study of \srcfull\ as part of GOALS. The high infrared luminosity of the pair results from the enhanced infrared emission from \src\ alone. \textit{Spitzer} found no or little evidence of significant Ne~[\textsc{v}] emission from Arp~302S, indicating no significant AGN activity in the southern galaxy of the pair. The Ne~[\textsc{v}]/Ne~[\textsc{ii}] and O~[\textsc{iv}]/Ne~[\textsc{ii}] line ratios of \src\ measured by \textit{Spitzer} indicates the presence of a weak AGN but contributing no more than 10--15\% of the infrared luminosity of \src, based on the scaling of local AGNs and starburst nuclei \citep{armus07}. A preliminary look at the \chandra\ observation of \srcfull\ found that \src\ shows significant X-ray emission, and AGN may dominate the hard X-ray emission band \citep{armus09}. \srcfull, consisting of two different types of galaxies within a gas-rich environment, is an excellent laboratory of a pair of interacting galaxies evolving along different paths.}

In this paper, we re-analyse the \chandra\ observation and conduct detailed spectral analysis and X-ray surface brightness study to search for extended X-ray emission by including another \xmm\ observation of \srcfull\ in the archive. \red{We aim to search for AGN X-ray activity in this infrared-luminous galaxy pair and investigate whether the measured soft X-ray luminosity is consistent with the SFR measurements at longer wavelengths. If the AGN activity is evident, we intend to identify the X-ray nature of the line-of-sight absorption, e.g., Compton-thin or Compton-thick, and study how its intrinsic X-ray luminosity compares to a typical Seyfert AGN.}


The paper is organised in the following way: section\,\ref{data} introduces our data reduction processes; section\,\ref{imag} presents an X-ray surface brightness study based on the \chandra\ observation; \red{section\,\ref{spec}} demonstrates the baseline model for the X-ray emission from \src; we discuss our results in section\,\ref{discuss}.

\section{Data Reduction} \label{data}

\subsection{\chandra}

The information of the two X-ray observations of \srcfull\ is in Tab\,\ref{tab_obs}. During the \chandra\ observation, the Advanced CCD Imaging Spectrometer (ACIS) was operated in the ACIS-S mode. Our data reduction for the \chandra\ observation was performed using CIAO v.4.15 and the calibration version 4.10.2 released on 15 November 2022. We reprocessed the data and extracted clean event lists using the \texttt{chandra\_repro} routine. We then extracted spectra using the SPECEXTRACT tool in CIAO. We specifically requested \texttt{correctpsf=yes} in SPECEXTRACT, and then the location is passed into the ARFCORR task. So, the X-ray count rates, especially those in Section\,\ref{imag}, have been corrected using the proper point spread functions. ACIS achieves a net count rate of $0.0169\pm0.0011$ cts s$^{-1}$, which is significantly lower than the count rate threshold for pile-up effects.

In this work, we focus on the galaxy \src\ of the pair, which is brighter than the other galaxy in the X-ray band. The source spectrum of \src\ was extracted from a circle of 20 arcsec centred at the centre of the galaxy. The background spectrum was extracted from a nearby, source-free circle of 60 arcsec. The spectrum is grouped using GRPPHA to have a minimum count of 20 per bin.

\subsection{\xmm}

We processed the \xmm\ observation of \srcfull\ using SAS v. 20.0 and the calibration version updated on 25 October 2022.

\subsubsection{EPIC}

The EPIC observations were operated in the Full Frame mode. We extracted a clean event list using the EMPROC and EPPROC tools respectively for MOS and pn observations. Good time intervals (GTIs) were selected by removing flaring particle background-dominated intervals, defined as periods when the single event (PATTERN=0) count rate in the >10\,keV band is higher than 0.35 cts s$^{-1}$ for MOS data and that in the 10--12\,keV band higher than 0.4 cts s$^{-1}$ for pn data. We extracted the source spectra from a circular region of 20 arcsec and the background spectra from a nearby, source-free region of 100 arcsec. Next, we used the RMFGEN and ARFGEN tools to create redistribution matrix files and auxiliary files. We group the spectra using GRPPHA to have a minimum count of 20 per bin. The net pn, MOS1 and MOS2 count rates of \src\ during our observation are $0.0399\pm0.0018$, $0.0099\pm0.0008$ and $0.0107\pm0.0008$ cts s$^{-1}$. They are lower than the X-ray count rate thresholds for the full-frame mode to have significant pile-up effects.

\subsubsection{OM}

\begin{figure}
    \centering
    \includegraphics[width=7cm]{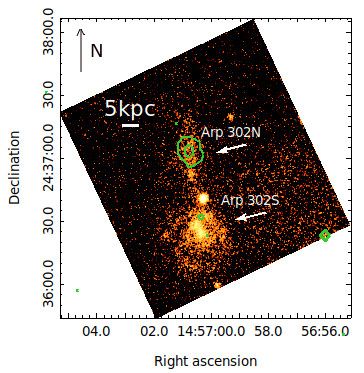}
    \caption{The OM image of the merging galaxies \srcfull\ in the UVW1 band overlaid with ACIS X-ray contours (0.2 and 0.6 counts per pixel). We find significant evidence of X-ray emission from Arp 302N and some from Arp 302S. The position angle of the edge-on galaxy \src\ is north-south.}
    \label{pic_om}
\end{figure}

The study of the UV and optical emission of \src\ is beyond the purpose of this work. Therefore, we show a UVW1 image of \src\ in Fig.\,\ref{pic_om} only for comparison with the X-ray emission region. The exposure of this image is 1\,ks. The image was extracted using the imaging mode data process tool OMICHAIN. 

\begin{table}
    \centering
    \begin{tabular}{cccc}
    \hline\hline
    Mission & ObsID & Date & Length \\
    \hline
    \xmm\   & 0670300101 & 2012-01-16 & 29\,ks \\
    \chandra & 7812 & 2006-12-17 & 15\,ks \\
    \hline\hline
    \end{tabular}
    \caption{The \xmm\ and \chandra\ observations analysed in this work.}
    \label{tab_obs}
\end{table}

\section{X-ray Flux Spatial Distribution} \label{imag}

Fig.\,\ref{pic_om} shows an OM  image of \srcfull\ in the UVW1 band. Two galaxies, marked by \src\ and Arp 302S, are clearly shown in the image. Two galaxies are separated by approximately 40 arcsec, which is around 26\,kpc at the distance\footnote{We assume $H_{0}$=67.8\,km\,s$^{-1}$\,Mpc$^{-1}$, $\Omega_{\rm matter}=0.308$, $\Omega_{\rm vacuum}=0.692$.} of the \srcfull\ (694\,pc per arcsec at $z=0.033$). The relatively large separation suggests that \srcfull\ is in an early phase of merging/interacting. 

The green contours in Fig.\,\ref{pic_om} show the X-ray emission from \srcfull\ measured by ACIS on \chandra. Some evidence of X-ray emission is found in Arp 302S. Most X-ray emissions originate in \src. The peak of the X-ray flux coincides with the centre of the galaxy. In this work, we focus on studying \src. 

\begin{figure}
    \centering
    \includegraphics[width=8cm]{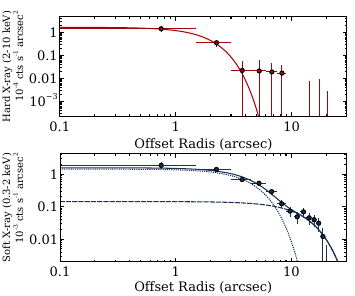}
    \caption{X-ray surface brightness of \src\ measured by ACIS. The centre is chosen to be the peak of the X-ray flux. The solid lines show the best-fit Gaussian distributions. The hard X-ray emission comes from a compact region with a radius of $\sigma<3.5$ arcsec (around 2\,kpc). The soft X-ray emission region is consistent with two Gaussian distributions of $\sigma=3.1\pm0.6$ arcsec (around 2\,kpc) and $\sigma=8.7^{+2.3}_{-1.7}$ arcsec (around 6\,kpc).}
    \label{pic_acis_flux}
\end{figure}

To further study the spatial distribution, we extract net X-ray count rate distribution for \src\ in 2--10\,keV and 0.3--2\,keV bands (see Fig.\,\ref{pic_acis_flux}). The background counts have been extracted using the same background region in Section \ref{data}. Thanks to the arcsec resolution of ACIS, we find that most hard X-ray emission comes from a compact region of 3 arcsec (around 2 kpc). The soft X-ray emission region is more extended (up to 18 arcsec, around 12 kpc). 

By modelling the distributions with Gaussian models, we obtain $\sigma$<2.4\,kpc of the hard X-ray emission region, which suggests most hard X-ray emission comes from close to the central nuclear region. Two Gaussian models are required to fit the soft X-ray distribution, one with $\sigma$=$2.1\pm0.4$\,kpc and the other with $\sigma$=$6.0^{+1.6}_{-1.2}$\,kpc. The two Gaussian distributions may suggest two distinct components in the 0.3--2\,keV band. We will discuss the soft X-ray spectral component of \src\ in the rest of the paper. The inner soft X-ray emission region has a similar size of a couple of kilo-parsec as the hard X-ray emission region. The outer soft X-ray emission region extends to more than 10\,kpc.

\section{Spectral Analysis} \label{spec}

\begin{figure}
    \centering
    \includegraphics[width=8cm]{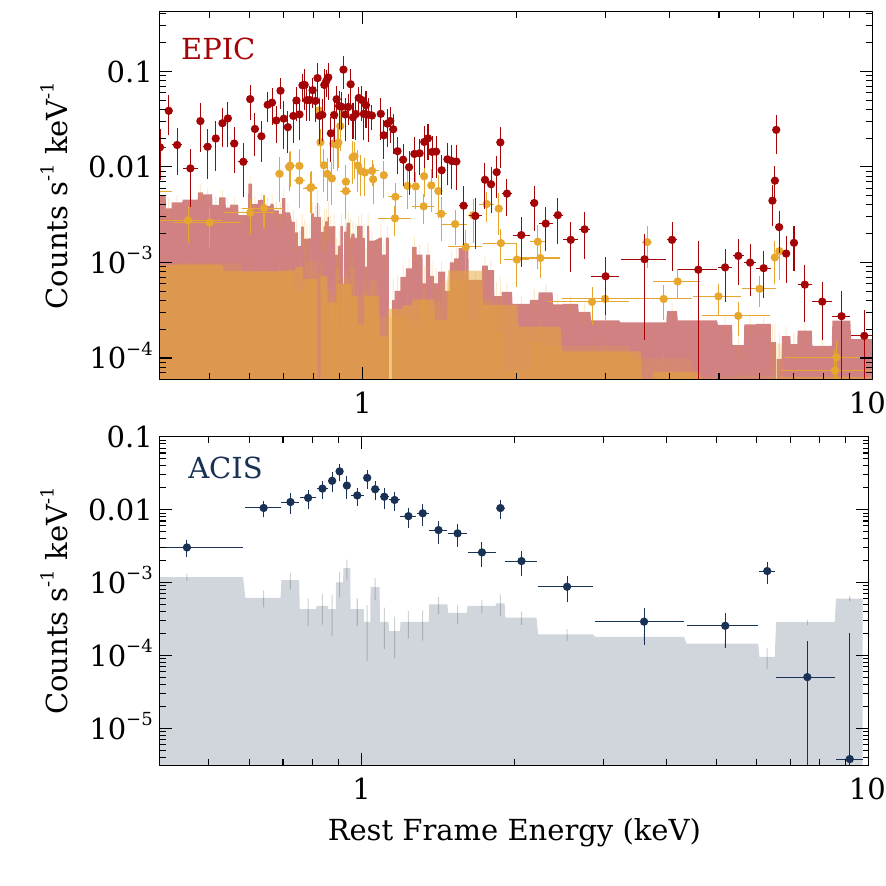}
    \caption{Source (crosses) and background spectra (shaded regions) of \src\ obtained by \xmm\ (Red: pn; orange: MOS1 and MOS2) and \chandra\ (black: ACIS). }
    \label{pic_data}
\end{figure}

\begin{figure}
    \centering
    \includegraphics[width=8cm]{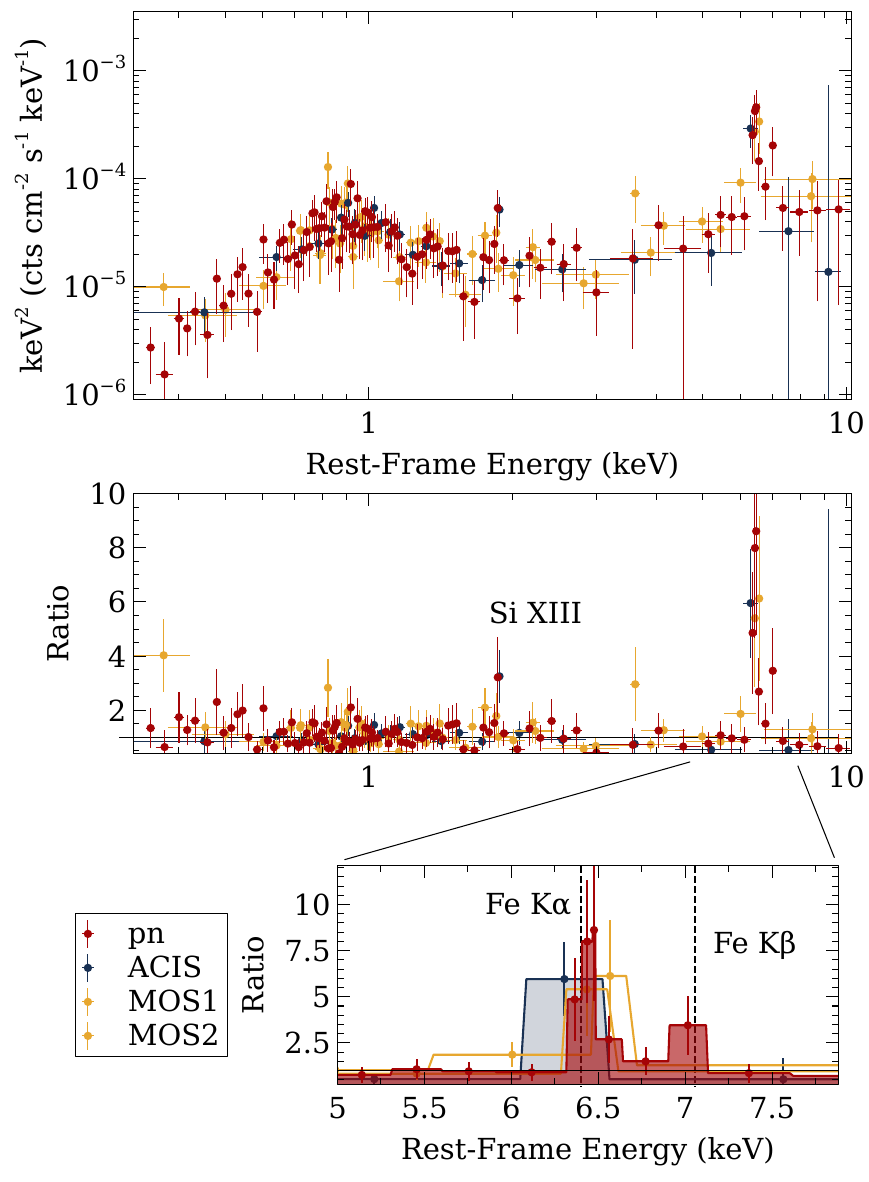}
    \caption{Top: folded spectra of \src\ but corrected for the effective area only for illustration purposes. Middle: data/model ratio plots for all four X-ray spectra using an absorbed continuum model including \texttt{zpowerlw} and \texttt{mekal}. We find evidence of Fe K$\alpha$, Fe K$\beta$ and Si \textsc{xiii} emission lines. Bottom: a zoom-in of the ratio plot in the iron emission band. The two vertical dashed lines mark the rest-frame energies of Fe K$\alpha$ (6.4\,keV) and K$\beta$ (7.06\,keV) emission. }
    \label{pic_uf}
\end{figure}

\subsection{Continuum Modelling}

We start the spectral analysis with continuum modelling. Fig.\,\ref{pic_data} and \,\ref{pic_uf} show an overview of the \chandra\ and \xmm\ spectra of \src. 


In the soft X-ray band, we model the hot gas emission using the \texttt{mekal} model in XSPEC. The \texttt{mekal} model calculates X-ray spectra from optically thin gas \citep{mewe85,mewe86,liedahl95,arnaud85,arnaud92}, e.g., in the star-forming region. We use the solar abundances calculated in \citet{wilms00} in the \texttt{mekal} model\footnote{In \citet{wilms00}, $Z_{\odot, \rm Fe}/ Z_{\odot, \rm H}$ equals $2.69\times10^{-5}$ and $Z_{\odot, \rm Si}/ Z_{\odot, \rm H}$ equals $1.86\times10^{-5}$. The silicon and iron abundances will be discussed in later sections}. The free parameters of \texttt{mekal} are the temperature of the gas $kT$ and the normalisation. Moreover, we follow the indication in \citet{armus09}, which shows similar spectral shapes as Fig.\,\ref{pic_uf}, by modelling the hard X-ray emission using the \texttt{zpowerlw} model. This power-law emission originates from the innermost accretion region of the AGN. The redshift parameters of \texttt{zpowerlw} and \texttt{mekal} are fixed at the source redshift. We also consider different line-of-sight absorption column densities for the \texttt{mekal} and \texttt{zpowerlw} models because they originate in two regions of the galaxy. The \texttt{ztbabs} model is used for this purpose. The redshift parameter of \texttt{ztbabs} is also fixed at the source redshift.

The \texttt{ztbabs} model is also used to account for Galactic absorption. \citet{willingale13} calculates the mean hydrogen column density $N_{\rm H, Gal}=N_{\rm HI}+N_{\rm H_{2}}=3.56\times10^{20}$\,cm$^{-2}$ in the direction of \src. Our X-ray data can not constrain the Galactic column density. This parameter is, therefore, fixed at this value. All \texttt{ztbabs} models consider the same abundances in \citet{wilms00}. An additional \texttt{constant} is used to account for cross-calibration uncertainty between different instruments and epochs. The \texttt{constant} value of pn is fixed at 1. The relative values for the other three spectra are MOS1/pn=$0.97\pm0.12$, MOS2/pn=$0.97\pm0.13$, ACIS/pn=$1.03\pm0.14$. They are all consistent with the calibration uncertainty \citep{madsen15}. These values are independent of the choice of the model. The ACIS/pn ratio is consistent with 1, suggesting no significant X-ray luminosity change in two observations separated by five years. 

In summary, the total continuum model is \texttt{constant * ztbabs * ( ztbabs1*mekal + ztbabs2*zpowerlw )} in XSPEC notations. This model describes the continuum emission of \src\ very well. Fig.\,\ref{pic_uf} shows the residuals in the continuum fitting. Three narrow emission lines are shown at 1.8, 6.4 and 7\,keV of the data. These three features are also shown in the folded spectra Fig.\,\ref{pic_data} and the top panel of Fig.\,\ref{pic_uf}. The 1.8\,keV emission feature was also shown in Fig.\,7 in \citet{armus09} where the same \chandra\ spectrum was considered. We will analyse these three emission lines and estimate their significance in a later section.

\subsection{The Metal Abundances of the Star-Forming Region}

The underabundance of Fe relative to $\alpha$ elements was found in nearby starbursts and local luminous infrared galaxies \citep{strickland04,grimes05,iwasawa11}. Some may be related to the enhanced production $\alpha$ elements in Type II SN \citep{martin02} or iron depletion in the boundary layer between the cold interstellar medium and SN winds \citep{strickland04}.

The CCD-resolution data of \src\ do not allow us to constrain the absolute abundances of elements in the \texttt{mekal} component. But by varying the iron abundance parameter while keeping a solar oxygen abundance, we may shed some light on the iron abundances in the star-forming region of \src. Because the $Z_{\rm Fe}/Z_{\rm O}$ value determines the Fe L/ O K emission ratio in the soft X-ray band. 

\begin{figure}
    \centering
    \includegraphics[width=7cm]{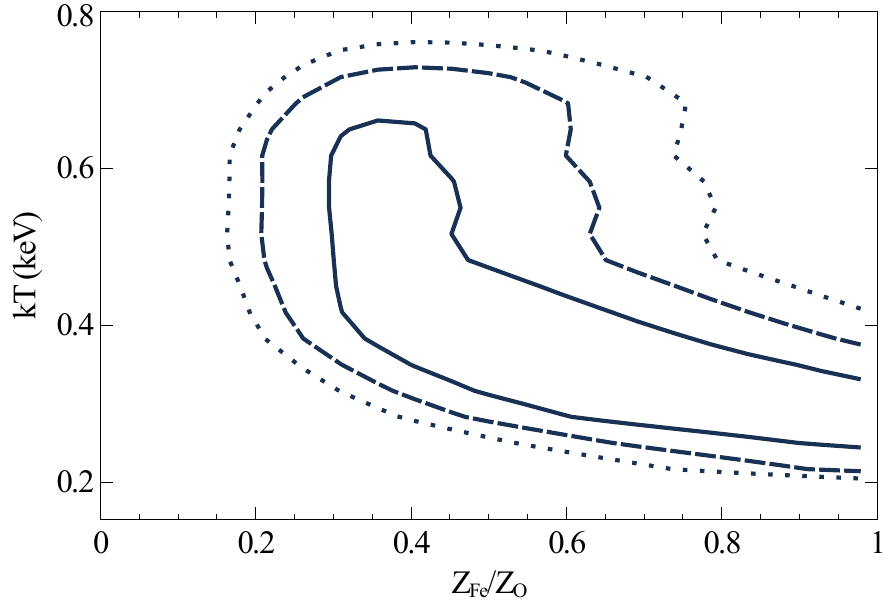}
    \caption{$\chi^{2}$ distribution on the relative iron abundance $Z_{\rm Fe}/Z_{\rm O}$ and the temperature of the gas. There is a degeneracy between these two parameters. The solid dashed and dotted curves show the 1, 2, 3$-\sigma$ uncertainty ranges.}
    \label{pic_zfe}
\end{figure}

In Fig.\,\ref{pic_zfe}, we show the measurement uncertainty of $Z_{\rm Fe}/Z_{\rm O}$ and $kT$ of the \texttt{mekal} model. There is clear evidence of degeneracy between the two parameters. Within the $1-\sigma$ uncertainty range, a model with $Z_{\rm Fe}=0.4Z_{\rm O}$ and $kT$=0.5\,keV can also explain the data. We, therefore, conclude that we cannot constrain the metal abundances of the hot gas of the star-forming region in \src. Although we measure the gas temperature by fixing the $Z_{\rm Fe}$ parameter of the \texttt{mekal} at solar in the following analysis, we note that a higher gas temperature is also possible if the metal abundances are lower.

\subsection{Detection of Narrow Emission Lines}

We fit the narrow emission lines shown in Fig.\,\ref{pic_uf} with Gaussian line models \texttt{zgauss} in XSPEC. The redshift parameter is fixed at the source redshift. The best-fit parameters are shown in Table\,\ref{tab_gauss}. The rest-frame energies of the three lines are $6.44\pm0.04$, $6.99\pm0.13$ and $1.85\pm0.04$\,keV, which align with the energies of Fe K$\alpha$, Fe K$\beta$ and Si\,\textsc{xiii} emission lines. Adding a Gaussian line model to fit each line decreases $\chi^{2}$ by 67.64, 17.31 and 10.55, respectively. Each Gaussian line model introduces two extra free parameters, energy and normalisation. The line width of the three line models is fixed at a small value $\sigma=10^{-3}$\,keV. If we free this parameter, we only obtain an upper limit\footnote{This upper limit corresponds to FWHM<0.236\,keV. For the 6.44\,keV line, the velocity dispersion is smaller than $v<0.236/6.4c=0.04c$.} of $\sigma<0.1$\,keV (90\% confidence level). We, therefore, fix this parameter in the model.

To estimate the significance of the lines, we conduct a null hypothesis test by simulating 1000 sets of spectral simulations. The same exposure time as in the archival MOS1, MOS2, pn and ACIS observations is considered. The simulation model assumes no additional lines at 6.44, 6.99 and 1.85\,keV. Only the best-fit continuum model shown in Table\,\ref{tab_gauss} (\texttt{constant * ztbabs * (ztbabs1*mekal + ztbabs2*zpowerlw)}) is used. We then search for emission line features by adding \texttt{zgauss} models with variable energy and normalisation parameters in the 1--3\,keV and 5--10\,keV and use the decrease of $\chi^{2}$ as an indication of goodness-of-fit improvement. The distribution of $\Delta\chi^{2}$ is shown in Fig.\,\ref{pic_sim}. 138 out of 1000 simulations have $\Delta\chi^{2}>10.55$ by adding a Gaussian line in the 1--3\,keV band of the simulated spectrum. We, therefore, estimate the significance of the 1.86\,keV emission line to be 1-138/1000=86.2\%. Similarly, the significance of the 6.99\,keV line is 98.6\%. No simulations have $\Delta\chi^{2}$ higher than 67.64 as for the observed Fe K$\alpha$ emission lines.

Narrow Fe\,K$\alpha$ emission lines are commonly seen in luminous infrared galaxies. AGN and star-forming regions, e.g., from \red{high-mass X-ray binaries (HMXBs) \citep[e.g.,][]{mineo12}},  may contribute to the observed Fe\,K$\alpha$ emission line. Based on the best-fit Gaussian line model in Table\,\ref{tab_gauss}, we estimate the flux of the Fe\,K$\alpha$ in \src\ is $2.5\times10^{-14}$\,\ergps, corresponding to a line luminosity of $7\times10^{40}$\,erg\,s$^{-1}$. Based on the  Fe\,K$\alpha$ line luminosity vs SFR relation compiled from observations of local luminous infrared galaxies \citep[$L^{\rm HMXB}_{\rm Fe K\alpha}$ (erg\,s$^{-1}$) $=(1.3\pm0.4)\times10^{37}$ SFR ($M_{\odot}$ yr$^{-1}$),][]{pereira11}, the strong Fe\,K$\alpha$ line in \src\ would suggest a very high SFR of 5400 $M_{\odot}$\,yr$^{-1}$, if the star-forming region dominates the Fe\,K$\alpha$ emission in \src. This conflicts with the study in the millimeter-wave observations of \src\, \citep[e.g., SFR=1-4 $M_{\odot}$ yr$^{-1}$,][]{lo97}. We conclude that the Fe\,K$\alpha$ emission line originates in the \red{AGN} rather than the galaxy's star-forming region.

\begin{table}
    \centering
    \begin{tabular}{ccccc}
    \hline\hline
    Models & Parameters & Units & Model 0 \\
    \hline
    \texttt{ztbabs1} & $N_{\rm H,1}$ & $10^{21}$\, cm$^{-2}$ & $6.0^{+1.3}_{-1.4}$ \\
    \texttt{mekal} & $kT$ & keV & $0.23^{+0.06}_{-0.02}$  \\
    & $Z_{\rm Si}$ & $Z_{\odot}$ & 1 (f)  \\
    & Norm1 & $10^{-4}$ & $5^{+4}_{-3}$  \\
    \hline
    \texttt{ztbabs2} &  $N_{\rm H,2}$ & $10^{20}$\, cm$^{-2}$ & $<5$ \\
    \texttt{zpowerlw} & $\Gamma$ & - & $1.4\pm0.2$  \\
    & Norm2 & $10^{-6}$ & $8.3\pm0.7$ \\
    \hline
    \texttt{zgauss} & Eline1 & keV & $6.44\pm0.04$ \\
                    & EW & eV & $3369^{+24}_{-52}$  \\
    \texttt{zgauss} & Eline2 & keV & $6.99\pm0.13$  \\
                    & EW & eV & $1125^{+14}_{-27}$  \\
    \texttt{zgauss} & Eline3 & keV & $1.86\pm0.04$\\
                    & EW & eV & $137^{+24}_{-13}$  \\
    \hline
     & $\chi^{2}/\nu$ & - & 192.47/196 \\
    \hline\hline
    \end{tabular}
    \caption{\red{A simple phenomenological model (Model 0) for the 0.5--10\,keV} spectra of \src. \red{The 2-10\,keV luminosity of the power-law component (\texttt{zpowerlw}) is $10^{41}$\,erg\,s$^{-1}$ based on the best-fit unabsorbed flux. This value is significantly lower than the expected X-ray luminosity by given measured infrared spectral measurements of the AGN  in \src\ using the scaling factors in \citep{spinoglio22}. See text for more details.}}
    \label{tab_gauss}
\end{table}

\begin{figure}
    \centering
    \includegraphics[width=8cm]{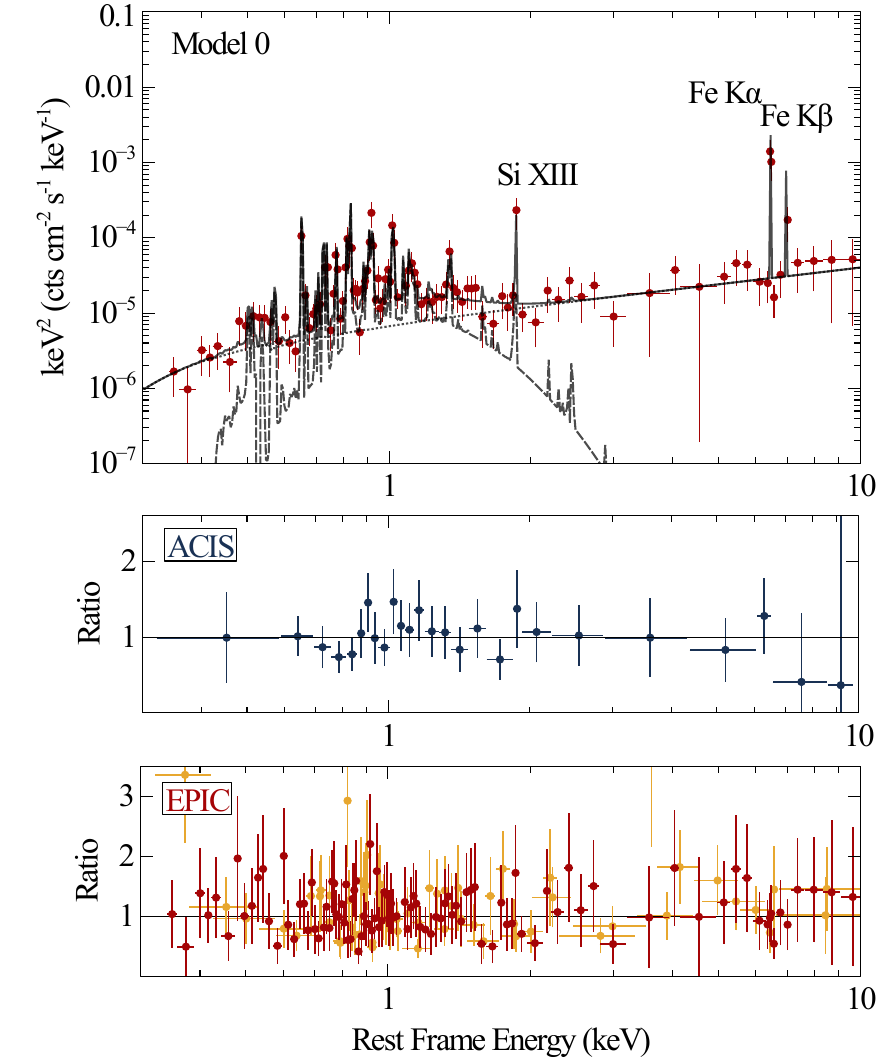}
    \caption{The best-fit Model 0 for the X-ray spectra of \src\ overlaid with the unfolded pn spectrum. Solid line: total model; dashed line: \texttt{mekal}. The detected emission lines in Fig.\,\ref{pic_uf} are modelled by Gaussian lines. The bottom two panels show corresponding data/model ratio plots.}
    \label{pic_gauss}
\end{figure}

\begin{figure}
    \centering
    \includegraphics[width=8cm]{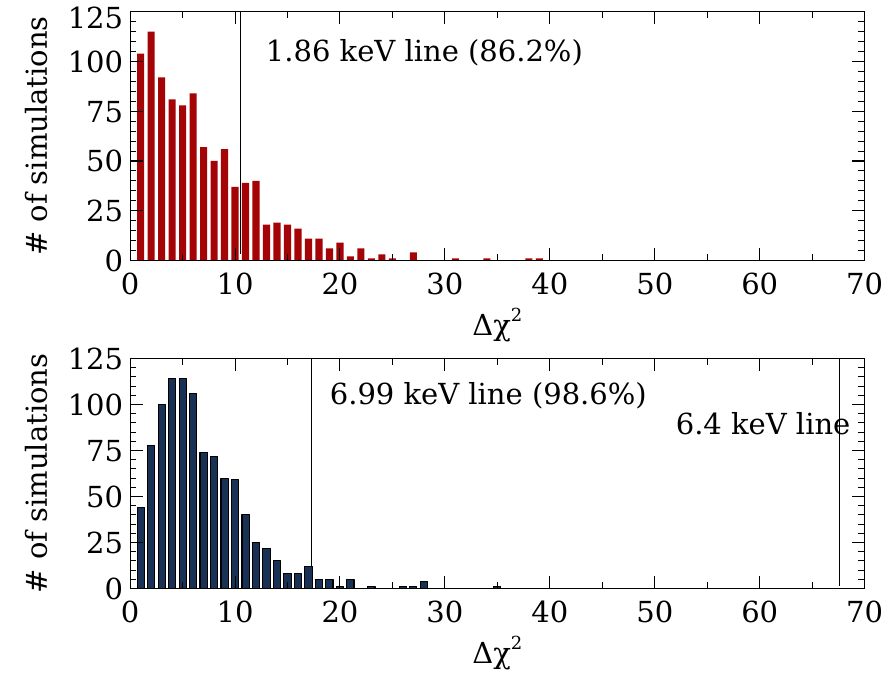}
    \caption{$\chi^{2}$ improvement by adding an additional Gaussian line in the 1--3\,keV and 5--10\,keV bands in 1000 simulations. See text for more details. 138/1000 simulations have $\Delta\chi^{2}>10.55$ in the 1--3\,keV band and 14/1000 simulations have $\Delta\chi^{2}>17.31$ in the 5--10\,keV band. The confidence of positive detection of Si\,\textsc{xiii} (1.86\,keV) and Fe\,\textsc{xxvi} (6.95\,keV) is respectively 86.2\% and 98.6\%.}
    \label{pic_sim}
\end{figure}

\subsection{\red{A Torus Model of the Obscured Power-Law Emission and Fe K Emission Lines}} \label{tori}

\red{In Model 0, we only obtain an upper limit for the line-of-sight column density of the power-law component (\texttt{zpowerlw}), $<5\times10^{20}$\,cm$^{-2}$. The little absorption in the continuum emission and the significant evidence of strong Fe K emission lines (equivalent width of >3000\,eV for Fe~K$\alpha$ and >1000\,eV for Fe~K$\beta$) forces us to re-evaluate the origin of the X-ray continuum emission.}

\red{More specifically, one can estimate the expected X-ray luminosity of \src\ by scaling the infrared measurements of the same object using statistical correlations drawn from a large sample of galaxies. \citet{armus09} measured the Ne~[\textsc{v}] (14.3~\textmu{m}), Ne~[\textsc{v}] (24.3~\textmu{m}) and O~[\textsc{iv}] (25.9~\textmu{m}) line flux of \src\ to be 1.2, 2.3 and 12.3 $\times10^{-17}$\,W\,m$^{-2}$ using \textit{Spitzer}. Using the relation compiled from one hundred nearby infrared galaxies in Fig.1--3 of \citet[][]{spinoglio22}, we expect the 2--10\,keV luminosity of \src\ to be around $10^{43}$\,erg\,s$^{-1}$, which is 100 times higher than the value of the best-fit power-law component in Model 0 ($10^{41}$\,erg\,s$^{-1}$ in the 2--10\,keV band calculated using the best-fit unabsorbed flux).}

\red{In this section, we explored the possible evidence of a heavily obscured power-law component in our spectral data below 10\,keV. The observed power-law component with little absorption in Model 0 may, instead, originate in scattering materials as seen in most Seyfert 2 AGN \citep[e.g.,][]{turner97} rather than direct emission from the hot corona.} 

\subsubsection{Model Set-Up}

\red{Following the indications of previous work on obscured AGNs in LIRGs \citep[e.g.,][]{yamada21,laloux22}, we explain the Fe~K$\alpha$ and K$\beta$ emission lines consistently with a torus model of gas and dust surrounding the AGN as in a typical Seyfert 2 AGN \citep{antonucci84}. We adopted the \texttt{mytorus} model \citep{murphy09}, where the covering factor of a toroidal-shaped torus is fixed at 0.5. The full model is \texttt{constant * ztbabs * (ztbabs1*mekal  + trans.obs * zpowerlw1 + scattered.con + fluor.lines + zpowerlw2)} (Model T) in XSPEC notation.  We used the coupled mode of the \texttt{mytorus} model, so the scaling factors of the Compton-scattered continuum emission (\texttt{scattered.con}) and fluorescent line emission (\texttt{fluor.lines}) are linked. Due to the lack of hard X-ray data, there is a degeneracy between the line-of-sight column density and the strength of the obscured, direct power-law emission (\texttt{zpowerlw1}). We, therefore, link the line-of-sight column density with the average column density of the torus. Such a model would allow us to consistently calculate the emission from the torus and line-of-sight absorption by giving the same column density.}

\red{In addition, Model 0 suggests a modest column density along our line of sight towards the collisionally ionised gas component \texttt{mekal} (around $6\times10^{21}$\,cm$^{-2}$). We, therefore, considered an independent line-of-sight absorption calculated by \texttt{ztbabs1} for the \texttt{mekal} model. We also allowed a variable Silicon abundance in the \texttt{mekal} model to fit the Si\,\textsc{xiii} emission line.}


\begin{table}
    \centering
    \begin{tabular}{cccccccc}
    \hline\hline
    Models & Parameters & Units &  Values\\
    \hline
    \texttt{ztbabs1} & $N_{\rm H,1}$ & $10^{21}$\, cm$^{-2}$ & $7.4^{+0.4}_{-0.3}$ \\
    \texttt{mekal1} & $kT_1$ & keV & $0.30\pm0.03$\\
                & $Z_{\rm Si}$ & $Z_{\odot}$ & $8\pm3$\\
                & Norm1 & $10^{-4}$ & $2.7^{+1.8}_{-0.3}$\\
    \hline
    \texttt{mytorus} & $\log(N_{\rm H,2})$ & cm$^{-2}$  & $>3\times10^{24}$\\
     \texttt{reprocessed}   &  $Z_{\rm Fe}$ & $Z_{\odot}$   & 1(f)\\
    & Norm2 & $10^{-5}$ & $91^{+3}_{-2}$\\ 
    \hline
    \texttt{obscured} & $N_{\rm H,3}$ & cm$^{-2}$  &$=N_{\rm H,2}$ \\
    \texttt{zpowerlw1} & $\Gamma_1$ & - & $=\Gamma_{1}$\\
                    & Norm3 & $10^{-5}$  &=Norm2\\
    \hline
    \texttt{scattered} & $\Gamma_2$ & - & $1.78^{+0.08}_{-0.09}$\\
    \texttt{zpowerlw2} & Norm4 & $10^{-5}$ &  $0.7\pm0.2$\\
  \hline
   & $\chi^{2}/\nu$ & - &  201.06/200\\
   \hline\hline
    \end{tabular}
    \caption{\red{Model T adopts the \texttt{mytorus} model to calculate Fe emission lines by given gas column density. The \texttt{mytorus} model has three components: reprocessed emission (continuum emission and lines), obscured direct power-law emission, and unobscured scattered power-law emission.}}
    \label{tab_torus}
\end{table}

\subsubsection{Results}

\red{Model T provides the spectra of \src\ a good fit with $\chi^{2}/\nu=201.06/200$. The best-fit parameters are shown in Table\,\ref{tab_torus}.}
\red{Assuming the line-of-sight column density ($N_{\rm H,3}$) is the same as the average column density of the torus ($N_{\rm H,2}$), we find a lower limit of these two parameters in Model T. Fig.\ref{pic_nh2} shows $\Delta\chi^{2}$ vs. $N_{\rm H,3,4}$. The 90\%-confidence lower limit is $3\times10^{24}$\,cm$^{-2}$, indicating Compton-thick medium. Note that the maximum of the allowed column density values in the \texttt{mytorus} model is $10^{25}$\,cm$^{-2}$. The scattering fraction of the continuum emission is Norm4/Norm2=$7\times10^{-6}/9\times10^{-4}=0.8\%$, consistent with the median value of known Sy2 AGN \citep{gupta21}.} In addition, to fit the observed Si~\textsc{xiii} emission, a super-solar silicon abundance of $8Z_{\odot}$ is required for the \texttt{mekal} model. 


\red{In summary, the observed unabsorbed power-law emission in \src\ is interpreted as the scattered emission from the circumnuclear materials in Model T. The direct power-law emission from the hot corona close to the SMBH is highly obscured in these models.} \red{This model} explains the lack of obscuration in the power-law continuum emission (\texttt{zpowerlw1}) based on a model similar to the ones for typical Seyfert 2 AGN. They naturally suggest a higher intrinsic AGN luminosity than Model 0. \red{See Section\,\ref{discuss} for more discussion regarding the intrinsic AGN luminosity.}

\begin{figure}
\centering
\includegraphics[width=8cm]{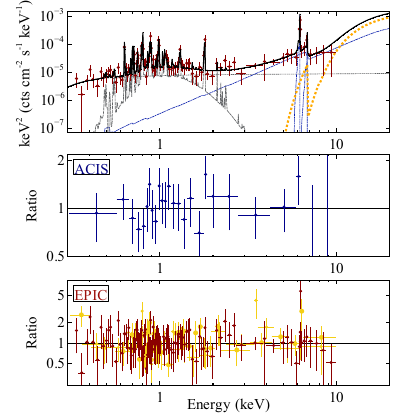}
\caption{\red{Top: best-fit \texttt{mytorus} models for the \xmm\ and \chandra\ spectra of \src. Black solid line: total model; grey dashed line: unobscured, scattered power-law emission; blue dash-dotted lines: Compton-scattered continuum and line emission components; orange dotted lines: obscured power-law components. The red crosses are the unfolded pn spectrum of \src. Middle: corresponding \chandra\ data/model ratio plot. Bottom: the same data/model ratio but for \xmm\ spectra (red: pn; yellow: MOS1 and 2).}}
\label{pic_mytorus}
\end{figure}

\begin{figure}
\centering
\includegraphics[width=8cm]{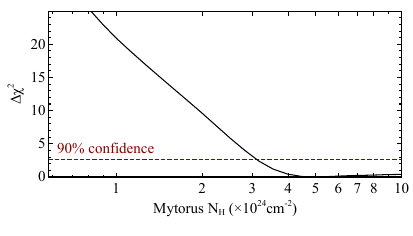}
\caption{\red{$\Delta\chi^{2}$ vs. the hydrogen column density of the torus in the \texttt{mytorus} model (Model T). The dashed line shows the 90\%-confidence lower limit is $3\times10^{24}$\,cm$^{-2}$.}}
\label{pic_nh2}
\end{figure}

\section{Discussion} \label{discuss}

\subsection{An X-ray view of the star-forming region in \src}

By analysing the soft X-ray surface brightness of \src\ measured by \chandra, we find that the soft X-ray emission of \src\ is consistent with two Gaussian distributions. The first component has a size of $\sigma=2$\,kpc; the second has a size of $\sigma=6$\,kpc. The first component has a similar emission size as the hard X-ray emission, while \chandra\ detected significant emission in the second component up to 12\,kpc. 

Detailed spectral modelling suggests that AGN and star-forming regions contribute to the soft X-ray emission of \src, \red{in agreement with the indication of two soft X-ray components in the image}. The more extended soft X-ray emission found in the surface brightness study likely corresponds to the star-forming component in the spectrum. We conclude that the size of the star-forming region extends to at least 12\,kpc in \src. \red{It is worthwhile to mention that extended soft X-ray emission, e.g., <2\,keV, spreading up to 2--10\,kpc is frequently found in other U/LIRGs, while hard X-ray emission, e.g., 2--10\,keV, is more compact within 1--3\,kpc \citep{iwasawa11}.} 

In addition to the soft X-ray emission region size in \src, we can also estimate the SFR of \src\ based on the relation between X-ray luminosity and SFR. Adapting the relation SFR$\approx L_{\rm 2-10keV}/10^{39} M_{\odot}$ yr$^{-1}$ \citep{franceschini03}, we estimate the SFR of \src\ to be 1--2\,$M_{\odot}$\,yr$^{-1}$ based on the best-fit \red{Model T}. This value is similar to the estimation in \citet{lo97} based on $L_{\rm IR}/M(\rm H_{2})$: \citet{lo97} argues that the high infrared luminosity of \src\ is not due to starbursts in the nuclear region, but its considerable amount of molecular gas, among which stars are forming at only SFR=1--4\,$M_{\odot}$ yr$^{-1}$.

\red{An intriguing result from our analysis is the evidence of the 1.86\,keV emission line in both \xmm\ and \chandra\ observations of \src\ separated by five years. Our confidence of a positive detection is around 86.2\,\% estimated from simulations. This line energy corresponds to a Si~\textsc{xiii} emission line. The diffuse gas in the star-forming region is the only explanation for this emission line in the soft X-ray band. A significant silicon overabundance of $Z_{\rm Si}=8Z_{\odot}$ is required to explain the data. The super-solar silicon abundance might be related to the past core-collapse supernovae in \src. For example, an over-abundance of silicon has been found in the circumstellar medium of Galactic core-collapse supernova remnants \citep[e.g.][]{bhalerao19}.}

\red{We explored the possibility of a second hot gas component (\texttt{mekal}), representing a separate star-forming region in \src, to explain the Si~\textsc{xiii} emission line without the requirement for a super-solar silicon abundance. Such a model needed two \texttt{mekal} components of $kT=0.33$ and 0.64\,keV. However, such a model would infer an overall X-ray luminosity of the star-forming region ten times higher than Model T, corresponding to an SFR of 20 $M_{\odot}$ yr$^{-1}$. This value does not match observations at longer wavelength \citep{lo97}. We, therefore, exclude such a model.}

The CCD-resolution data from \xmm\ and \chandra\ cannot \red{resolve the detailed spectral lines of the hot gas in the star-forming region}, but future microcalorimeter-resolution data, e.g. from \textit{Athena} \citep{nandra13}, will do. We show simulations of one X-IFU observation based on Model T. The evidence of the finest emission lines, such as C\,\textsc{iv} emission, will answer this question. \textit{AXIS} \citep{mushotzky19} is a future X-ray mission with CCD-resolution data but a much higher effective area than \chandra. Despite no detailed lines resolved in the soft X-ray band, e.g., <1\,keV, \textit{AXIS} will \red{still} show a significant advantage in detecting  Fe\,K$\alpha$ and K$\beta$ emission over \chandra. 







\subsection{The AGN activity in \src}


\red{Our \xmm\ and \chandra\ spectra of \src\ show that a power-law component dominating the <10\,keV band has little absorption. Assuming this power-law emission originates from the central accretion region of the AGN, e.g., the hot coronal region near the SMBH, we estimate the AGN  of \src\ has a 2--10\,keV luminosity of $10^{41}$\,erg\,s$^{-1}$. This value contrasts the expected X-ray luminosity scaled by previous \textit{Spizter} measurement of the AGN activity in \src\ \citep{armus09} using a scaling correlation between infrared neon and oxygen emission line luminosity and X-ray luminosity in \citet{spinoglio22}. See Section\,\ref{tori} for more details.}  

\red{We, therefore, proposed a model with a heavily obscured power-law emission in the data. Our analysis fits the AGN emission with the \texttt{mytorus} model following the indications of previous work on other obscured AGNs in LIRGs \citep[e.g.,][]{yamada21,laloux22}. Such a model allows us to consistently calculate Fe K emission lines and photoelectric absorption by assuming the same value for the torus's line-of-sight and average column density.}

When both a direct, obscured power law and a scattered, unobscured power law are considered, the inferred intrinsic AGN X-ray luminosity of \src\ is higher in Model \red{T} than in Model 0. The resulting 2--10\,keV unabsorbed luminosity of the power-law component is $1.0\times10^{43}$\,erg\,s$^{-1}$. \red{This value matches the expectation based on previous \textit{Spitzer} measurements of the AGN activity of this galaxy \citep{armus09}.} Assuming a BH mass of $10^{8}M_{\odot}$, this luminosity corresponds to $L_{\rm 2-10keV}/L_{\rm Edd}=10^{-3}$. This value is reasonable for a \red{typical Seyfert 2 AGN \citep{bian07}}. \red{In addition, the scattering fraction of the power-law emission of \src\ is around 0.7\% (Norm4/Norm2), similar to other Seyfert 2 AGNs \citep{gupta21}. Although no report yet on the optical classification of \src, e.g., from SDSS,} \src\ might host a Seyfert 2 AGN where the X-ray emission from the accretion region is highly obscured. 

\red{A column density of at least $3\times10^{24}$\,cm$^{-2}$ is required to explain the apparent lack of obscured power-law emission below 10\,keV and the evidence of significant Fe K emission lines, suggesting a Compton-thick environment in the AGN of \src. Our spectral analysis adds more supporting evidence of hidden AGN activity in the sample of early merging galaxies. The Compton-thick circumnuclear environment revealed by our X-ray spectral analysis provides a complementary view of the dust and gas in the LIRG \src.} 

\begin{figure*}
    \centering
    \includegraphics[width=17cm]{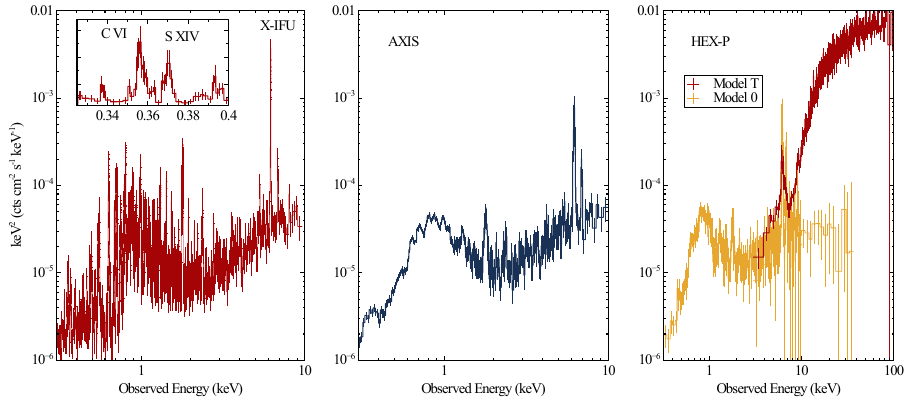}
    \caption{Simulated \textit{Athena} X-IFU, \textit{AXIS} and \textit{HEX-P} spectra of \src\ using \red{Model T}. An exposure of 100\,ks is considered for all three simulations. Left: the evidence of ionised emission lines, such as C\,\textsc{iv}, holds the answer to the origin of the soft X-ray emission below 0.7\,keV in \src. \textit{Athena} will provide an unprecedented constraint on these emission lines. Middle: \textit{AXIS} has a CCD resolution detector. Although it cannot resolve all the emission lines in the soft X-ray band, it shows significant improvement in measuring the  Fe\,K$\alpha$ and Fe\,K$\beta$ emission of the photoionised gas compared to \chandra. It is important to note the arcsecond angular resolution of \textit{AXIS}. Combined with the high effective area, \textit{AXIS} will more precisely measure the surface brightness of \src\ in X-rays. Right: \textit{HEX-P} will provide an important view of the hard X-ray emission from \src. Future hard X-ray observations will provide direct evidence of an obscured AGN in \src, based on which we can study the accretion state of the AGN, e.g., a low-luminosity AGN or a typical Seyfert 2 AGN. A simulated low-energy detector spectrum based on only \red{Model 0} is shown in this panel for clarity.}
    \label{pic_spec_sim}
\end{figure*}

\red{Our best-fit torus model predicts a 14--195\,keV flux of $5.4\times10^{-12}$\,\ergps\ from \src, which is lower than the sensitivity of the Swift-BAT survey\footnote{\red{The 105-month Swift-BAT survey is a uniform hard X-ray all-sky survey with a sensitivity of $8.4\times10^{-12}$\,\ergps\ over 90\% of the sky in the 14--195\,keV band \citep{oh18}.}}. Due to the lack of hard X-ray data, we had to assume the line-of-sight column density and the average column density of the torus are identical in Model T. However, this assumption may not hold. For example, evidence of the clumpiness in the torus has been found in both X-ray and infrared observations \citep[e.g.,][]{krolik88,mason09,laha20,zhao21}.} Future high-sensitivity, hard X-ray missions like \textit{HEX-P} \citep{madsen19} will more precisely measure the obscured power-law component \red{in the hard X-ray band of \src\ and test for different torus models. In Fig.\,\ref{pic_spec_sim}, we show two simulations, one based on Model 0 without obscured power law-shaped AGN emission and one based on Model T with a full torus model. \textit{HEX-P} will easily distinguish two models in the hard X-ray band. More relevant simulations for the \textit{HEX-P} spectra of obscured AGN can be found in \citet{boorman23}.}


\section{Conclusion}

We present an analysis of the \chandra\ and \xmm\ observations of the non-starburst, luminous infrared galaxy \src. \src\ is one of the two galaxies in \srcfull. The two galaxies are separated by 26\,kpc, suggesting an early galaxy interaction/merging phase. Previous observations at longer wavelengths found that the high infrared luminosity of \src\ is not due to a high SFR but a large amount of molecular gas spreading up to 23\,kpc. By studying its X-ray emission, we made the following conclusions:

\begin{itemize}
    \item \chandra\ detects significant soft X-ray emission (0.3--2\,keV) up to 12\,kpc in \src\ from the peak of the X-ray emission. By modelling the X-ray surface brightness, we find two Gaussian-like emission regions,  one with $\sigma$=$2.1\pm0.4$\,kpc and the other with $\sigma$=$6.0^{+1.6}_{-1.2}$\,kpc. The smaller soft X-ray emission region has a similar size as the hard X-ray emission (2--10\,keV).
    \item We detect significant Fe K$\alpha$ and Fe K$\beta$ (98.6\% confidence) emission lines from \src. \red{We explore the possibility of consistently explaining the Fe K emission lines and line-of-sight absorption by given column density. Such a model suggests that a Compton-thick circumnuclear environment with $N_{\rm H}>3\times10^{24}$\,cm$^{-2}$. Future hard X-ray telescopes, such as \textit{HEX-P}, will be able to more precisely measure the torus geometry, test for clumpiness in the torus and determine its hard X-ray luminosity.}
    \item \red{Our best-fit X-ray model suggests an intrinsic X-ray (2--10\,keV) luminosity of $10^{43}$\,erg\,s$^{-1}$, consistent with the previous \textit{Spitzer} measurements of the AGN activity in \src\ \citep{armus09}. The unobscured power-law emission with a luminosity of $10^{41}$\,erg\,s$^{-1}$ that dominates the <10\,keV band is the Thomson scattered emission from the distant Compton-thin circumnuclear materials. The scattering fraction is 0.7\% in \src. The line-of-sight column density, scattering fraction and intrinsic X-ray luminosity suggests a Seyfert 2-like AGN in \src. Future optical spectral observations will provide further confirmation of the AGN classification.}
    \item Both the \red{unobscured, scattered power-law emission} and the \red{hot gas} in the star-forming region in \src\ contribute to the soft X-ray emission of \src. We estimate \src\ has SFR=$1-2 M_{\odot}$\,yr$^{-1}$ using the X-ray luminosity of the hot gas in the star-forming region. This result matches previous CO measurements at longer wavelengths \citep{lo97}. 
    \item We detect a Si~\textsc{xiii} emission line in the ACIS and pn spectra of \src\ (86.2\% confidence). It can be explained by the super-solar silicon abundance of the hot gas in the star-forming region and associated with past core-collapse supernovae in \src. Similar \red{silicon} enrichment was observed in the circumstellar medium of Galactic supernova remnants. \red{Future high-resolution X-ray spectral data, e.g., from \textit{Athena}, will uniquely probe the ionisation state and metallicity of the hot gas in the star-forming regions in the LIRGs with obscured AGNs.}
\end{itemize}

\section*{Acknowledgements}

J.J. acknowledges support from the Leverhulme Trust, the Isaac Newton Trust and St Edmund's College, University of Cambridge. W.B. acknowledge support by the Science and Technology Facilities Council. We acknowledge the valuable discussion on plasma modelling with Timothy R. Kallman and Honghui Liu. 

\section*{Data Availability}

The \xmm\ and \chandra\ data are available for download at https://heasarc.gsfc.nasa.gov. The \texttt{mytorus} model is available at https://www.mytorus.com. No new data or models were produced in this work.




\bibliographystyle{mnras}
\bibliography{maxi1535.bib} 

\bsp	
\label{lastpage}
\end{document}